%Paper: hep-th/9309023
%From: VENEZIA@crnvma.cern.ch
%Date: Fri, 03 Sep 93 17:46:08 SET

\magnification=1200
\hsize 15true cm \hoffset=0.5true cm
\vsize 23true cm
\baselineskip=20pt

\font\grande=cmr10 scaled \magstep4
\font\medio=cmr10 scaled \magstep2
\outer\def\beginsection#1\par{\medbreak\bigskip
      \message{#1}\leftline{\bf#1}\nobreak\medskip\vskip-\parskip
      \noindent}

\def \me {\buildrel <\over \sim}
\def \Me {\buildrel >\over \sim}
\def \pa {\partial}
\def \ra {\rightarrow}

\def \fb {\overline \phi}
\def \rb {\overline \rho}
\def \pb {\overline p}
\def \pr {\prime}
\def \se {\prime \prime}
\def \ti {\tilde}

\def \b {\beta}
\def \a {\alpha}
\def \ap {\alpha^{\prime}}

\def \ga {\gamma}
\def \sg {\sigma}

\def \da {\delta}
\def \ep {\epsilon}
\def \r {\rho}

\def \Om {\Omega}
\def \noi {\noindent}

\def\sqr#1#2{{\vcenter{\hrule height.#2pt\hbox{\vrule width.#2pt
height#1pt \kern#1pt\vrule width.#2pt}\hrule height.#2pt}}}

\def\lsim{\mathrel{\rlap{\lower4pt\hbox{\hskip1pt$\sim$}}
    \raise1pt\hbox{$<$}}}         %less than or approx. symbol
\def\gsim{\mathrel{\rlap{\lower4pt\hbox{\hskip1pt$\sim$}}
    \raise1pt\hbox{$>$}}}         %greater than or approx. symbol

\nopagenumbers
\line{\hfil CERN-TH.6955/93}

\vskip 3.3 true cm

\centerline{\grande   Inflation, Deflation, and Frame-Independence}
\bigskip
\centerline{\grande in String Cosmology}

\vskip 1.8true cm
\baselineskip=15pt
\centerline{M. Gasperini \footnote{*}
{Permanent address: {\it Dipartimento di
Fisica Teorica, Via P. Giuria 1, 10125 Turin, Italy}}
and G. Veneziano}
\baselineskip=20pt
\centerline {\it Theory Division, CERN, Geneva, Switzerland}

\vskip 2 true cm

\baselineskip=15pt
\centerline{\medio Abstract}
\noi
The inflationary scenarios suggested by the duality properties of
string cosmology  in the Brans-Dicke (or String) frame
are shown to correspond to accelerated contraction (deflation) when
 Weyl-transformed to the Einstein frame. We point out that the basic
virtues of inflation (solving the flatness and  horizon
problems, amplifying vacuum fluctuations, etc.) have physically
equivalent counterparts in the deflationary (Einstein-frame) picture.
This could be the answer to some objections recently raised to
superstring cosmology.

\vskip 2 true cm
\noi
{CERN-TH.6955/93}\par
\noi
{July 1993}\par

\vfill\eject

\footline={\hss\rm\folio\hss}

\pageno=1

{\bf 1. Introduction}

A potential source of difficulty for extended-inflation
models [1]  based on a Brans-Dicke theory of gravity [2]  is the choice
of the correct frame (metric) in which to describe the space-time
 geometry at a cosmological level. One may wonder,
  in particular, in which frame the
metric should be of the inflationary type, and satisfy the conditions
required to avoid the problems of the standard cosmological scenario.

 While the choice of
  the Einstein (E) frame (in which the Einstein-Hilbert term takes the
General-Relativity form) usually simplifies calculations and
is quite popular, there are physical motivations for choosing instead
 the Brans-Dicke (BD) frame, in which matter couples to the metric-tensor
in the standard way [3].
 Arguments in favour of the BD choice can also be given
 in string theory [4], where the BD frame metric coincides
 with  the $\sg$-model metric to which test strings are
directly coupled. Thus free string motions follow geodesic surfaces with
respect to the BD (not the E) metric.

The physical observable properties of a given model should be
independent, of course, from the field redefinition (Weyl rescaling)
connecting BD and E frames.
 And indeed, in the case of extended inflation, the metric describing a
phase of power-law inflation (with variable Newton constant)
 in the BD frame,
 is transformed into a metric, which is still describing power
inflation (of the slow-roll type, with exponential potential)
 in the E frame,
 as discussed for instance in [5].

In a string theory context, the role of the BD scalar is played by the
dilaton field. In such case, as pointed out in [6],
 there appear to be serious difficulties in arranging a
successful phase of dilaton-driven, power-law, extended inflation, at
least if theoretically motivated dilaton potentials are used.
 On the other hand, the cosmological equations obtained
  from the low-energy
string effective action show that the dilaton can drive (even in the
absence of a potential) a phase of accelerated expansion. This phase,
 supposedly describing the Universe before the big-bang
 (so-called ``pre-big-bang'' [7]), is characterized by being just the
  ``dual'' counterpart (in the sense of ref. [8])
  of the  ``post-big-bang'' standard cosmology.

The ``pre-big-bang'' phase corresponds, in the BD frame, to a
superinflationary expansion. When transformed to the E frame, however,
the same metric describes, as we shall see, a contracting Universe.
  Apparently, this represents
  a difficulty for the whole scenario, since the presence or
absence of inflation (and of its bonuses) would seem to
 become frame-dependent.

In this paper we shall show that, on the contrary, even in the E frame
the solutions of the string cosmology equations provide an adequate
description of the inflationary phase, provided we generically mean,
by ``inflation'', a phase of cosmological evolution that is able to avoid
the problems (see for instance [9]) related to the decelerated
kinematics of the standard cosmological model.

At the same time, and irrespectively of strings and/or   BD theory,
 we shall argue that the solution of many of
  the standard-cosmology problems
 achieved by inflation is also possible through
the introduction of an early phase of accelerated contraction,
 that we shall
call deflation. This will be the content of the following section.

\vskip .5 cm
{\bf 2. Inflation vs. deflation}
\smallskip

It is well known that there are three possible classes of inflationary
evolution [10], corresponding to a curvature scale
that is constant (De Sitter inflation), decreasing (power
inflation) or increasing (superinflation). Less known, however, seems to
be the fact that in a phase of growing curvature the solution of
the standard cosmological puzzles can be realized in two ways, namely by
a metric describing $\it {either}$ accelerated expansion,
 $\dot a >0, \ddot a >0$,
$\it {or}$ accelerated contraction, $\dot a <0, \ddot a <0$
($a$
is the scale factor of a homogeneous and isotropic model, and a dot
denotes differentiation with respect to cosmic time).

A possible equivalence of superinflation
and accelerated contraction is clearly pointed out by an elementary
analysis of the so-called flatness problem. If we want
 the contribution of the spatial curvature $k$ to be
suppressed with respect to the other terms of the cosmological
equations, then the ratio
$$
r_1={k\over a^2H^2} ={k\over \dot {a}^2}~~,
 ~~ H \equiv \dot a /a \; , \eqno(2.1)
$$
  must tend to zero during the inflationary era. Such a
condition is clearly satisfied by a metric that behaves, for $t \ra
+\infty$, as
$$
a\sim t^{\a}~~~~,~~~~t>0~~~~,~~~~\a > 1\; , \eqno(2.2)
$$
but also by a metric, which, for $t\ra 0_-$,  behaves as
$$
a\sim (-t)^{\b}~~~~,~~~~t<0~~~~,~~~~\b < 1 \; . \eqno(2.3)
$$

 The case (2.2) corresponds to power inflation,
and includes the standard De Sitter exponential inflation
 in the limit $\a \ra \infty$.
The second case, (2.3), corresponds, for $\b <0$, to
 the well-known case of
pole inflation (superinflationary expansion, $\dot a , \ddot a , \dot H$
all positive). For $0< \b <1$ it describes instead an
accelerated contraction, or deflation
 ($\dot a , \ddot a , \dot H$ all negative). In
both cases the curvature scale is growing, and $H, \dot H$ diverge as $
t \ra 0_-$.

A  deflationary phase   (2.3), with $0<\b<1$, may also
provide a solution to the so-called horizon problem.
The presently observed large-scale homogeneity and isotropy
requires the proper
size of the particle horizon to become  large enough during the
inflationary era, and  to go  to infinity in the limiting case in which
inflation extends for ever in the past. This means that the
integral
$$
d_p(t)= a(t)\int_{t_1}^{t} dt^{\pr} a^{-1}(t^{\pr}) \eqno(2.4)
$$
must diverge, if $a$ is the inflationary scale factor, when $t_1$
approaches the maximal past extension of the cosmic time coordinate for
the given cosmological manifold.

For the metric (2.3) such a limiting time is $-\infty$, and $d_p \ra
\infty $ for $t_1 \ra -\infty$, so that there are no particle horizons
in a phase of accelerated contraction.

As a consequence of accelerated contraction,  causally
connected regions are pushed out of the event horizon, just as in the
standard inflationary expansion. It is true that the proper size of   a
causally connected region tends to contract, asymptotically, like the
scale factor. For a patch of initial size $d_1\sim (-t_1)$ one finds
in fact, from eqs. (2.3) and (2.4), that $d_p \ra[a(t)/a(t_1)]d_1$ for
$|t|<<|t_1|$. However, the proper size of the event horizon, defined by
$$
d_e(t)= a(t)\int_{t}^{t_2} dt^{\pr} a^{-1}(t^{\pr}) \eqno(2.5)
$$
($t_2$ is the maximal allowed future extension of the cosmic time
coordinate), contracts always faster than $d_p$. Indeed, $t_2=0$ for the
metric (2.3), and one finds that $d_e(t) \sim (-t)$ for $t \ra 0$. The
ratio of the two proper sizes at small $t$
$$
r_2(t) = {d_p(t)\over d_e(t)}  \sim (-t)^{\b -1} \eqno(2.6)
$$
shows that the causally connected regions will always cross the horizon,
asymptotically, not only in the case of superinflationary expansion ($\b
<0$), but even in the deflationary case ($0<\b<1$ ).

We note, for later convenience, that the conditions for a successful
resolution of the horizon and flatness problem, when expressed in terms of
the conformal time coordinate $\eta$ ($a= dt/d\eta$), are
exactly the same   for both superinflationary expansion and accelerated
contraction. Moreover, if the contracting phase is long enough to solve
the horizon problem, then also the flatness problem is automatically
solved (and vice versa),  as in standard inflation.

Indeed the ratio $r_2$ scales in conformal time like $\eta^{-1}$, while
the ratio $r_1$ scales like $\eta^2$. The horizon problem is solved if
$r_2(\eta_f)$, evaluated at the end of the accelerated evolution
($\eta=\eta_f$), is larger than the present value $r_2(\eta_0)\simeq
1$, rescaled down at $\eta_f$. This implies
$$
{|\eta_i|\over |\eta_f|} \Me {|\eta_0|\over |\eta_f|} \simeq 10^2
({T_{rh}\over eV}) \; . \eqno(2.7)
$$
Here $\eta_i$ denotes the beginning of the contracting (or expanding)
accelerated evolution, $T_{rh}$ the final reheating temperature at
$\eta=\eta_f$, and the last equality holds in the hypothesis of
standard, adiabatic, radiation-dominated and matter-dominated expansion
from $\eta_f$ down to the present time $\eta_0$.

The solution of the flatness problem, on the other hand, is obtained if
the ratio $r_1$ at the end of the accelerated phase is tuned to a
value that is small enough, so that the subsequent decelerated evolution
leads to a present value of $r_1$ satisfying the condition
$r_1(\eta_0)\me 1$. This means
$$
({\eta_f\over \eta_i})^2 \me ({\eta_f\over \eta_0})^2 \;  , \eqno(2.8)
$$
which is clearly equivalent to eq. (2.7), and which implies a resolution
of the flatness and horizon problems (as well as of their rephrasing in
terms of the entropy [9]) for both
expanding and contracting metrics of the type (2.3).

Besides solving the kinematical problems, a phase of successful
inflation is also expected to  efficiently amplify the vacuum
fluctuations of the metric background. We shall conclude this section by
noting that such an amplification can also be provided
  by a long period of
deflation.

Consider, for instance, the amplification of tensor perturbations
$h_\mu^\nu$ (similar arguments hold for the scalar case also). In a
four-dimensional conformally flat background, the wave equation for each
Fourier component of $h$ can be written in terms of the rescaled
variable $\psi=ah$ as [11]
$$
\psi^{\se} +(k^2-{a^{\se}\over a})\psi=0 \eqno(2.9)
$$
(a prime denotes differentiation with respect to conformal time). In a
realistic case, the phase of accelerated evolution is followed by the
standard radiation-dominated expansion, with $a\sim \eta$, and the
amplification of the fluctuations can be described as a process of
graviton production from the vacuum (such an approach will be used in
Section 3). Equivalently, in a Schr\"odinger-like language, the process
corresponds to a parametric amplification of the
perturbation wave function [11], which is oscillating at $\eta \ra
\pm \infty$, and evolves with a power-law behaviour in the regions where
the co-moving frequency $k$ is negligible with respect to the effective
potential $a^{\se} /a$ of eq. (2.9).

By inserting into (2.9) a generic parametrization (in conformal time) of
the accelerated metric, $a(\eta)= (-\eta)^{-\da}$, one finds indeed that
the solution behaves like
$$
h \sim A_{\pm} {e^{\pm ik\eta}\over a} ~~~~~,~~~~~ k\eta>>1 \eqno (2.10)
$$
$$
h \sim A + B {(-\eta)\over a^2} = A+B(-\eta)^{1+2\da}~~~,~~~ k\eta<<1
\eqno(2.11)
$$
($A_{\pm}, A, B$ are integration constants). In the case of accelerated
expansion ($\da >0, a\ra \infty$ for $\eta \ra 0_-$), the perturbations
are amplified because their amplitude tends to stay constant in the
$\eta \ra 0$ limit, instead of decreasing adiabatically as in the
oscillating regime (2.10).

In the case of deflation ($\da <0, a \ra 0$ for $\eta \ra
0_-$), the amplification process is even more efficient than in the
previous case, as the amplitude of $h$ grows (with respect to the
adiabatic red-shift of the subsequent radiation-dominated expansion)
even in the oscillating regime. Moreover, as shown by eq. (2.11), $h$ may
even  grow asymptotically (instead of being constant)
 provided $\da < - 1/2 $.
 As we shall see in Section 3,  this condition is
satisfied  in particular, in the E frame, by a 3-dimensional
phase driven by stretched strings.

Note that the amplification coefficient
corresponding to a phase of accelerated contraction is different, in
general, from the one corresponding to a phase of accelerated expansion.
It is just because of this difference that the perturbation spectrum may
remain unchanged, when an inflationary background is transformed
into a deflationary one through a conformal rescaling, as we shall see in
the following Section.
\vskip .5 cm
{\bf 3. Pre-big-bang cosmology in the Brans-Dicke and Einstein frames}
 \smallskip
In a string cosmology context [7,12], a global (at least
semi-quantitative) description of the evolution and symmetries of the
early Universe is expected to be provided by the low-energy string
effective action, possibly supplemented by the action $S_m$ for
macroscopic matter sources:
$$
S= -{1\over 16 \pi G}\int d^{d+1}x \sqrt{|g|} e^{-\phi}
[R +(\pa_\mu \phi)^2-{1\over 12}H^2_{\mu\nu\a} +V] + S_m \eqno(3.1)
$$
Here $H_{\mu\nu\a}$ is the antisymmetric tensor field strength, and $V$
a (possibly non-zero) dilaton potential.

In this paper we will consider a $(d+1)$-dimensional,
anisotropic   metric background of the Bianchi I type, with time-dependent
dilaton,
$$
g_{00}=1~~~,~~~g_{ij}=-a_i^2\da_{ij}~~~,~~~\phi=\phi (t)
{}~~~,~~~i,j=1,2,...,d \eqno(3.2)
$$
and with vanishing $H_{\mu\nu\a}$ and $V(\phi)$. The additional matter
sources, which are decoupled from the dilaton in this frame,
 will be represented by a perfect fluid with anisotropic pressure:
 $$
T_0^0= \r ~~~,~~~ T_i^j=-p_i\da_i^j =- \ga_i \r \da_i^j \; . \eqno (3.3)
$$
By defining as usual [8,7,12]
$$
\fb =\phi - \ln \sqrt{|g|} ~~~,~~~ \rb = \r \sqrt{|g|}~~~,~~~
\pb = p\sqrt{|g|} \eqno(3.4)
$$
the field equations following from the variation of the action (3.1) can
be written in the form [8]
$$
\dot {\fb} ^2- 2 \ddot {\fb} + \sum_i H_i^2 =0 \eqno(3.5)
$$
$$
\dot {\fb} ^2 -\sum_i H_i^2 = \rb e^{\fb} \eqno(3.6)
$$
$$
2(\dot H_i- H_i\dot {\fb}) =\pb _i e^{\fb} \eqno(3.7)
$$
where $H_i=\dot a_i/a_i$, and we use units in which $8\pi G =1$. Their
combination gives the usual conservation equation
$$
\dot {\rb} + \sum_iH_i \pb _i =0 \; . \eqno(3.8)
$$

By applying the general procedure illustrated in [7],
the background field
variables can be separated, and the equations can be integrated exactly,
by introducing a suitable time-like coordinate $x$ such that
$$
\rb ={1\over L}{dx\over dt} \eqno(3.9)
$$
($L$ is a constant with dimensions of length, in such a way that $x$ is
dimensionless). For constant $\ga_i$ we
obtain  the following general exact solution of eqs.(3.5--3.7) (a similar
 problem was first solved  in a different context  in [13]):
$$
a_i= a_{0i}|(x-x_+)(x-x_-)|^{\ga_i/\a} |{x-x_+ \over x-x_-}|^{\a_i}
\eqno(3.10)
$$
$$
e^{\fb} = e^{\phi_0}
|(x-x_+)(x-x_-)|^{-1/\a} |{x-x_+ \over x-x_-}|^{-\sigma}
\eqno(3.11)
$$
$$
\rb = {\a \over 4 L^2} e^{\phi_0}
|(x-x_+)(x-x_-)|^{(\a - 1) /\a} |{x-x_+
 \over x-x_-}|^{-\sigma} \eqno(3.12)
$$
where
$$
\a =1- \sum_i \ga_i^2~~,~~ \sigma = \sum_i\a_i
\ga_i ~~,~~
\a_i = {\a x_i+\ga_i(\sum_i \ga_i x_i -x_0)\over
\a [(\sum_i\ga_ix_i-x_0)^2+\a(\sum_i x_i^2-x_0^2)]^{1/2}}
$$
$$
x_{\pm}= {1\over \a} \{\sum_i\ga_i x_i -x_0 \pm
[(\sum_i\ga_ix_i-x_0)^2+\a(\sum_i x_i^2-x_0^2)]^{1/2} \}
\eqno(3.13)
$$
and $a_0, \phi_0, x_0, x_i$ are integration constants.

This solution has various interesting properties, which we shall discuss
elsewhere [14]. Here we only note that there
 are two curvature singularities
at $x=x_{\pm}$, and that the region between the singularities is
unphysical, in the sense that the critical density parameter
$$
\Om (x) \equiv {\r e^{\phi} \over (d-1) \sum_i H_i^2} =
{(x+x_0)^2- \sum_i(\ga_i x + x_i)^2 \over (d-1) \sum_i
(\ga_i x +x_i)^2} \eqno(3.14)
$$
becomes negative. This parameter tends to zero at the singularities, and
in this limit the metric (3.10) goes over to the vacuum solutions
  of string cosmology [15,8]. For $x \ra x_{\pm}$ one finds indeed
$$
a_i(t) \sim |t-t_{\pm}|^{\b_i^{\pm}}~,~ \eqno(3.15)
$$
where
$$
\b_i^{\pm}= {x_i \pm \ga_i x_{\pm} \over x_0 + x_{\pm}}
{}~~~~~,~~~~~ \sum_i(\b_i^{\pm})^2 =1 ~.~\eqno(3.16)
$$

However, because of the neglect in the original action (3.1) of truly
``stringy'' contributions (such as $\ap$ and loop  corrections), this
solution is not expected  to provide a reliable description of
the very high curvature regime. The appropriate range of validity of the
solution is
instead the large $|x|$ limit, and in particular $x \ra -\infty$,
where it provides a typical example of pre-big-bang evolution,
characterized by acceleration and growing curvature scale [7].

If we consider, in particular, the isotropic case with negative pressure
($a_i=a, \ga_i=\ga <0$ for all $d$ spatial directions), then at large
negative $x$ we have $|x|\sim |t|^{\a /(2- \a )}$, and the solution
(3.10-3.12) becomes, in this limit,
$$
a(t) \sim (-t)^{2\ga /(1+d\ga^2)}~~~~,~~~~ \fb(t) \sim -{1\over \ga}
\ln a
$$
$$
\phi = \fb + d \ln a \sim {d\ga -1\over \ga} \ln a~~~~,
{}~~~~ \rb \sim a ^{-d\ga}\; . \eqno(3.17)
$$
For $\ga =-1/d$, which is the typical equation of state for a perfect gas
of stretched (or unstable) strings [16], one thus recovers the
particular solution already considered in [7,12] (``string-driven''
pre-big-bang). More generally, however, the background (3.17) describes
a phase of superinflationary expansion, $H>0, \ddot a/a>0$, and growing
curvature scale, $\dot H>0$, for all $\ga<0$.

This is the picture in the BD frame, which may be regarded as the natural
one in a string theory context [3]. The passage to the E frame, defined
as the frame in which the graviton and dilaton kinetic terms are
diagonalized  and  the action takes the standard   form,
$$
S_E= {1\over 16 \pi G}\int d^{d+1}x \sqrt{|\ti g|}
[-\ti R(\ti g)+{1\over 2}\ti{g}^{\mu\nu}\pa_\mu \ti \phi
\pa_\nu \ti \phi ] + S_m ~, \eqno(3.18)
$$
is obtained through the conformal rescaling
$$
\ti{g}_{\mu\nu} =g_{\mu\nu} e^{-2\phi /(d-1)}~~~~,~~~~
\ti \phi =\sqrt{{2\over d-1}}  \phi .
\eqno(3.19)
$$
The E-transformed scale factor, $\ti a$, and cosmic time coordinate,
$\ti t$, are thus related to the original BD ones by
$$
\ti a = a e^{-\phi/(d-1)}~~~~~,~~~~~ d\ti t =dt e^{-\phi/(d-1)} .
\eqno(3.20)
$$

The pre-big-bang configuration (3.17) becomes, in the
E frame,
$$
\ti a (\ti t) \sim (- \ti t)^\b~~~~,~~~~
\ti \phi \sim \sqrt{{2\over d-1}}{(d-1)(1-d\ga) \over (\ga -1)} \ln \ti
a
$$
$$
\ti \r \sim \ti a^{-2/\b}~~~~~,~~~~~ \b=
{2(1-\ga)\over (d-1)(1+d\ga^2) -2(d\ga-1)}
\eqno(3.21)
$$
where $\ti \r$ is conformally related to the original density $\r$ as
$$
\ti \r = \r {\sqrt{|g|}\over \sqrt{|\ti g|}} = \r e^{\phi (d+1)/(d-1)}
\eqno(3.22)
$$
(see for instance [17]). For
all $d >1$ and $\ga <0$, the transformed metric (3.21) satisfies
$$
{\ddot {\ti a} \over \ti a} <0 ~~~~~,~~~~~\ti H <0~~~~~,~~~~~ \dot {\ti H}
<0 \; , \eqno(3.23)
$$
where $\ti H =\dot {\ti a} /\ti a$, and the dot denotes here
differentiation with respect to $\ti t$. The BD superinflation thus
becomes an accelerated contraction of the type (2.3).

This result is a consequence of the non-trivial evolution of the dilaton
background that determines the transformation between the two frames,
and it is  of crucial importance. It implies that, if inflation is
long enough in the BD frame to solve the kinematical problems of the
standard model, then such problems are also  solved in the E frame.
 Indeed,
according to eq. (3.20), the two frames have the same conformal time
$$
d\ti \eta ={d\ti t \over \ti a (\ti t)}={dt\over a(t)} =d\eta
\eqno(3.24)
$$
and we have shown in Section 2 that the conditions to be satisfied for
solving the kinematical problems, when expressed in conformal time, are
the same for both superinflationary expansion and accelerated
contraction.

Moreover, the spectrum of the metric perturbations amplified in the
course of the background evolution is also the same in both frames. This
can be easily shown by considering , for instance, the case of tensor
perturbations, and assuming a generic model of background evolution
characterized by the transition (at $\eta=\eta_1$) from the accelerated
phase to the standard radiation-dominated one. In conformal time, such
evolution can be parametrized as
$$
a\sim (-\eta)^{-\da}~~~~,~~~~\phi \sim \ep \ln a~~~~,~~~~
\eta<< -\eta_1
$$
$$
a\sim \eta ~~~~,~~~~ \phi \sim const ~~~~,~~~~ \eta >>-\eta_1 \; .
\eqno(3.25)
$$

In order to verify the equality of the spectral behaviour, it is crucial
to take into account the fact that not only the background solutions,
but also the perturbation equations are different, when the frame is
changed. In the BD frame, the tensor perturbation equation contains
explicitly the contribution of the dilaton background, and for each
component of $h_\mu^\nu$ the equation can be written [7,17]
$$
\psi^{\se} +(k^2-V)\psi=0 \; , \eqno(3.26)
$$
where [7,17]
$$
\psi = h a^{(d-1)/2}e^{-\phi/2}
$$
$$
V={(d-1) a^{\se}\over 2 a} -{\phi^{\se}\over 2}
+{(d-1)(d-3)a^{\pr 2}\over 4
a^2}+{\phi^{\pr 2}\over 4}-{(d-1)a^{\pr}\phi^{\pr}\over 2 a} \; .
\eqno(3.27)
$$

By matching the solutions of (3.26) corresponding to the two phases of
background evolution, one can compute the Bogoliubov coefficients
relating $|in\rangle$ and $|out \rangle$ vacua, and describing the
associated graviton production. For co-moving frequencies $k$
that are small
enough with respect to the height of the effective potential barrier
($k\eta_1 <<1$), the modulus of the Bogoliubov coefficient is [7,18]
$$
|c_-(k)|\simeq (k\eta_1)^{-|\nu|-1/2} \eqno(3.28)
$$
where
$$
\nu={\da\over 2}(d-1-\ep) +{1\over 2} \eqno(3.29)
$$
and the corresponding spectral distribution of gravitons is determined
as $\r (k)= k^4|c_-|^2$. In the case of four-dimensional exponential
inflation ($\da =1,d=3,\ep =0$) one thus finds, in particular, the flat
Harrison-Zeldovich spectrum.

In the more general case of the background (3.17), one finds that, in
conformal time, the kinematics is parametrized according to eq. (3.25) by
$$
\da=-{2\ga \over 1-2\ga +d\ga^2}~~~~,~~~~ \ep= {d\ga -1 \over \ga}\; .
\eqno(3.30)
$$
The coefficient $|\nu|$ determining the pre-big-bang graviton spectrum
in the BD frame is thus
$$
|\nu| ={1\over 2}|{d\ga^2-1 \over 1-2\ga +d\ga^2}| \; . \eqno(3.31)
$$

In the E frame, there is no explicit dilaton contribution to the
perturbation equation for $h$, which is exactly the same equation as
that satisfied by a minimally coupled scalar field [11] (the dilaton
contribution, however, is implicitly contained in the rescaled
metric background). Such an equation can still be written in the form
(3.26), (3.27), but with $\phi=const$. As a consequence, the spectral
coefficient $|\nu|$ of eq. (3.28) is determined by the metric background
only, and becomes
$$
\nu={\ti \da \over 2}(d-1) +{1\over 2} \; ,
\eqno(3.32)
$$
where $\ti \da$ is the exponent parametrizing, in conformal time, the
evolution of the contracting E metric (3.21):
$$
\ti \da = {2(\ga-1)\over (d-1)(1-2\ga +d\ga^2)} \; .  \eqno(3.33)
$$
This value, when inserted into eq. (3.32), provides exactly the same
expression for $|\nu|$ as in eq. (3.31), and thus the same graviton
spectrum as in the BD frame.

We want to stress, finally, that the same results hold in the case of
conformal vacuum backgrounds, namely for solutions of eqs. (3.5-3.7) with
$\r = p =0$ [8,15] (the general vacuum solution for the action (3.1)
with non-zero $H_{\mu\nu\a}$ is given in [19]).

In the vacuum case the analogous of the isotropic, $d$-dimensional
solution (3.17) is, in the BD frame,
$$
\eqalign{
a_{\mp}(t)&\sim |t|^{\mp 1/\sqrt{d}} \cr
\phi_{\mp}(t)&\sim -(1\pm \sqrt{d})\ln |t|=
\pm(\sqrt{d} \pm d)\ln a_{\mp} \cr} \eqno(3.34)
$$
The two signs correspond to the two duality-related solutions [8], and
the upper sign describes a ``dilaton-driven'', pre-big-bang,
superinflationary expansion for $t$ ranging from $-\infty$ to $0$.

In the E frame the solution (3.34) becomes (in conformal time)
$$
\ti a (\ti \eta) = |\ti \eta|^{1/(d-1)}~~~,~~~~
\ti \phi (\ti \eta) = \mp \sqrt{2d(d-1)} \ln \ti a \eqno(3.35)
$$
and it always describes an accelerated contraction of the type (2.3),
independently of the choice of sign in eq. (3.34). It is interesting to
note that the duality transformation, which is represented in the BD
frame as an inversion of the scale factor and a related dilaton shift,
$$
a_+ \ra a_-=a_+^{-1} ~~~,~~~ \phi_+ \ra \phi_-= \phi_+-2d \ln a_+
\eqno(3.36)
$$
  becomes, in the E frame, a transformation between what we may call
a   strong-coupling
and a weak-coupling regime, $\ti \phi \ra -\ti \phi$, without changing
the metric background described by $\ti a$.
\vskip .5 cm
{\bf 4. Conclusions}
 \smallskip
The main goal of this paper has been to show that, for what concerns the
solution of the kinematical problems (horizon, flatness) of the
standard model, and the amplification of the vacuum fluctuations, an
accelerated contraction of the metric is equally good as an accelerated
expansion.

This observation was motivated by the fact (also discussed in this
paper) that accelerated contraction is the behaviour of the metric in a
general pre-big-bang cosmological string scenario, when seen in
the Einstein frame. Indeed, as already stressed in [7], there are only
two ways of implementing a phase of cosmic acceleration and simultaneous
growth of the curvature scale: accelerated contraction and
superinflationary (or pole-like) expansion. The latter corresponds to
the pre-big-bang picture in the conformally related Brans-Dicke frame.

Obviously, a contracting phase cannot dilute the relic abundance of
some unwanted remnants, such as the monopoles of the GUT phase
transition. However, the same is true for the pre-big-bang scenario in
the BD frame, as well as for all models in which the phase of inflationary
expansion occurs at some higher fundamental (near  Planckian)
  scale, which is indeed what is expected in a string
cosmology context. In this respect, we recall
[7] that a pre-big-bang phase should be regarded not necessarily as
an alternative, but possibly as a complement to the more conventional
inflationary models, which cannot be extended (at least  semiclassically)
  beyond the Planck era.

  Moreover, it is clear that   deflationary contraction is
  adiabatic for what concerns radiation, just like the usual
inflationary expansion. Therefore, as recently stressed also in [20], a
kinematical modification of the standard model can explain the
large present value of the cosmic black-body entropy, only if the
accelerated evolution is matched to the standard one through a phase
dominated by some non-adiabatic process (the so-called ``reheating''
era).

In the BD picture of the pre-big-bang scenario (see eq. (3.17)), the
radiation is supercooled and diluted with respect to the sources that
drive   inflation. The conservation equation (3.8) leads in fact to an
effective source temperature $T_s \sim a^{-d\ga}$, which
 grows together with
the scale factor for $\ga <0$, and satisfies
$$
{T_s \over T_r}={\r _\ga \over \r_r} \sim a^{1-d\ga} \eqno(4.1)
$$
($r$ corresponds here to the radiation-like equation of state,
$\ga=1/d$). The reheating process is thus expected to represent, in this
frame, a sort of non-adiabatic conversion of the hot sources into
radiation, such as a possible isothermal decay of the highly excited
states of a gas of stretched strings [7].

In the E frame (see eq. (3.21)) the fluid sources satisfy a modified
conservation equation,
$$
\dot{\ti \r}+d \ti H(\ti \r +\ti p)-{\dot{\ti \phi } \over
\sqrt{2(d-1)}} (\ti \r -d \ti p) =0 \; .\eqno(4.2)
$$
Radiation  still evolves adiabatically, now with a
 blue-shifted temperature  because of the contraction, $\ti T _r
\sim \ti a ^{-1}$. The effective temperature of the pre-big-bang sources
is also blue-shifted, however, since, in the perfect
 fluid approximation,
eq. (4.2) leads to
$$
\ti T_s \sim \ti a^{(d^2\ga^2+1-d\ga-d\ga^2)/(\ga-1)} \eqno (4.3)
$$
and thus
$$
{\ti T_s \over \ti T_r}={\ti \r _s \over \ti \r_r} \sim
\ti a^{\ga (d-1)(d\ga -1)/(\ga -1)} \; . \eqno(4.4)
$$
For $\ti a \ra 0$ the temperature of the sources that drive
the acceleration ($\ga <0, d>1$)
is always growing, even with respect to the radiation
temperature. The
physical picture of   reheating as a non-adiabatic decay of the hot
sources is still valid, therefore, also in the Einstein frame.

We would like to stress, finally, that the absence of problems related
to some ``preferred frame'' description of a string cosmology inflation
is to be ascribed, to a large extent, to the crucial role
 played by the dilaton field,
which transforms conformally a superinflationary expansion into
a deflationary
contraction. This is to be traced back to the duality properties of the
string effective action [8,12,19,21],
and   thus gives support to the consistency of an
approach to string cosmology based on the
effective action (3.1).

\vskip 1 cm
\noi
{\bf Acknowledgements}

\noi
We are grateful to N. Sanchez for useful
discussions. One of us (G.V.) wishes to thank A. De R\'ujula and E.W. Kolb
for raising questions that motivated in part this investigation.

\vfill\eject
\centerline {\bf References}
 \smallskip
\item{1.}D. La and P.J. Steinhardt, Phys. Rev. Lett. 62 (1989) 376.

\item{2.}C. Brans and C.H. Dicke, Phys. Rev. 124 (1961) 925.

\item{3.} See e.g. Ya.B. Zeldovich and I.D. Novikov,
 ``The Structure and the

Evolution of
the Universe'',  University of Chicago Press (1982).

\item{4.}N. Sanchez and G. Veneziano, Nucl. Phys. B333 (1990) 253;

L.J. Garay and J. Garcia-Bellido, Nucl. Phys. B400 (1993) 416.

\item{5.} E.W. Kolb, D.S. Salopek and M.S. Turner, Phys. Rev. D42 (1990)
 3925;

E.W. Kolb, Physica Scripta T36 (1991) 199.

\item{6.}B.A. Campbell, A. Linde and K.A. Olive, Nucl. Phys. B335
(1991) 146;

 R. Brustein and P.J. Steinhardt, Phys. Lett. B302 (1993) 196.

\item{7.}M. Gasperini and G. Veneziano,
Astropart.  Phys. 1 (1993) 317.

\item{8.}G. Veneziano, Phys. Lett. B265 (1991) 287.

\item{9.}A. H. Guth, Phys. Rev. D23 (1981) 347.

\item{10.}F. Lucchin and S. Matarrese, Phys. Lett. B164 (1985) 282.

\item{11.}L.P. Grishchuk, Sov. Phys. JETP 40 (1974) 409.

\item{12.}M. Gasperini and G. Veneziano, Phys. Lett. B277 (1992) 256.

\item{13.}V.A. Ruban and A.M. Finkelstein, Gen. Rel. Grav. 6 (1975)
601.

\item{14.}M. Gasperini, N. Sanchez and G. Veneziano, in preparation.

\item{15.}M. Mueller, Nucl. Phys. B337 (1990) 37.

\item{16.}M. Gasperini, N. Sanchez and G. Veneziano, Nucl. Phys. B364
(1991) 365.

\item{17.}S. Kalara, N. Kaloper and K.A. Olive, Nucl. Phys. B341 (1990)
252.

\item{18.}M. Gasperini and M. Giovannini, Phys. Rev. D47 (1993) 1529.

 \item{19.}K.A. Meissner and G. Veneziano, Mod. Phys. Lett. A6 (1991)
3397.

\item{20.}Y. Hu, M.S. Turner and E.J. Weinberg, FERMILAB-Pub-92/363-A.

\item{21.}A.A. Tseytlin, Mod. Phys. Lett. A6 (1991) 1721;

A. Sen, Phys. Lett. B271 (1991) 295;

M. Gasperini, J. Maharana and G. Veneziano, Phys. Lett. B272
(1991) 277;

ibid. B296 (1992) 51;

S.F. Hassan and A. Sen, Nucl. Phys. B375 (1992) 103;

A.A. Tseytlin and C. Vafa, Nucl. Phys. B372 (1992) 443;

A.A. Tseytlin, Class. Quantum Grav. 9 (1992) 979;

E. Kiritsis, ``Exact duality symmetries in CFT and string theory'',

 CERN-TH.6797/93.

\end